\documentclass[epsf,amstex]{article}
\usepackage [dvips]{graphicx}
\usepackage{amsmath}
\usepackage{times}\usepackage{french}
\usepackage{psfig}
\usepackage{epsfig}
\usepackage{amssymb}

\newcommand{\dd}{\text{d}}

\newcommand{\ee}{\text{e}}

\newcommand{\p}{\partial}

\begin{document}

\newenvironment{dessins}{{\bf Figures}}{}
\begin{center}
{\huge{\bf Field-theory for reaction-diffusion processes \\
with hard-core particles}}
\end{center}
\vspace{2cm}
\begin{center}
{Fr\'ed\'eric van Wijland$^{(a)}$}
\end{center}

\noindent {\small $^{(a)}$Laboratoire de Physique Th\'eorique, Universit\'e de Paris-Sud,
91405 Orsay cedex, France.}\\\\

\begin{center}{\bf Abstract}\\
\end{center}
{\small We show how to build up a systematic bosonic field-theory for a general reaction-diffusion process involving 
{\it hard-core} particles in
arbitrary dimension. We criticize a recent approach
proposed by Park, Kim and Park ({\it Phys. Rev. E} {\bf 62} (2000)). As a testbench for our method, we show how
to recover the equivalence between asymmetric diffusion of excluding particles
and the noisy Burgers equation.}

\vskip 2cm

Pr\'epublication L.P.T. Orsay 00/75.
\newpage
\section{Introduction}
A general formalism to describe nonequilibrium statistical physics is still
lacking, but recent progress has been achieved on the important issue of the description of collective phenomena in such systems
(phase transitions in nonequilibrium steady-states~\cite{dickman}, emergence of long-range correlations
in driven systems~\cite{ziaschmittmann}, cellular automata~\cite{droz}, self-organized
criticality~\cite{alexvrevue}). Indeed steps
forward have been made in the last two decades in that
direction by exploiting the formal analogy of their field-theoretic formulation with
static critical phenomena. Of course, numerical approaches and exact solutions
have played an equally important r\^ole, but this communication will
deliberately ignore those aspects. The beauty of the field-theoretic
formulation --when available-- is that, combined with a renormalization group analysis, it provides the theorist with a systematic
analytic tool for the calculation of physical observables in the scaling regimes of
interest. However, a vast category of reaction-diffusion
processes, cellular automata, driven lattice gases and other related stochastic
models fail to be exactly mapped onto continuous field theories. This
difficulty is due to the fact that the integer degrees of freedom (often particle
numbers) usually have an integer upper limit which prevents from exploiting the
familiar mapping first introduced by Doi~\cite{doi} and recently revived by Cardy~\cite{cardyrevue}. For instance, in a reaction-diffusion
process with mutually excluding particles, local particle numbers are 0 or 1,
but one might imagine other processes with other similar constraints ({\it e.g.} the
contact threshold transfer process defined by Rossi {\it et al.}~\cite{alexv} in which a given site is occupied by no more than two particles). In some instances, the exclusion
constraint can be phenomenologically accounted for, as done {\it e.g.} by Zia and
Schmittmann~\cite{ziaschmittmann} when they build up effective Langevin equation to describe the
dynamics of driven-diffusive systems. Very often the evolution operator can be
exactly mapped on a quantum spin-chain. But it is only in one space
dimension that (in favorable cases) one can exploit the toolbox of integrable
systems~\cite{schutz} (local particle numbers usually restricted to 0 and 1, little being
known~\cite{malte} on spin-1 (and higher) chains for stochastic problems). \\ 

A coveted goal is therefore to be able to build up a systematic and exact field-theoretic path-integral
formulation that can, at least formally, account for limitations in local particle
numbers.\\

Several attempts have been made to incorporate the exclusion constraint
into a field-theory of standard type. As already mentioned, one may achieve this goal by
phenomenologically exploiting the physical knowledge of the system (Zia and
Schmittmann~\cite{ziaschmittmann}
or Cardy~\cite{cardyijmp}), but this
is by no means exact, and can in general only be implemented safely in high
space-dimension. Directly in dimension 1, other approaches exist, such  as that
proposed by Cardy~\cite{cardy} and Brunel {\it al.}~\cite{vkf} or by Mobilia and
Bares~\cite{mobiliabares}. In the former the authors succeeded in constructing a
field-theory in a systematic way, but of fermionic type, which unfortunately proved, from a
technical view, rather difficult to analyze, while the latter, though efficient
in the pair annihilation reaction, seems difficult to extend to other processes.\\

In what follows we shall present the derivation of the path-integral formulation
for systems of mutually excluding particles of a single species. Where
appropriate we will also indicate how to extend the theory to cope with several species or other
constraints on local particle numbers. We shall illustrate how the mapping works
on the case of asymmetric diffusion, and how one can recover the equivalence
with the noisy Burgers equation.

\section{Hard-core particles using a bosonic formulation}
\subsection{Master equation and bosonic formalism}
The evolution of a configuration $n\equiv \{n_i\}$ of local particle numbers
is encoded in a master equation for the probability $P(n,t)$ to
observe configuration $n$ at time $t$. The master equation for
$P(n,t)$ is equivalent to an evolution equation for the state vector
$|\Psi(t)\rangle=\sum_n P(n,t)|n\rangle$, which we write in the form
\begin{equation}
\p_t|\Psi\rangle=-\hat{H}|\Psi\rangle
\end{equation}
The operator $\hat{H}$, which acts on the space spanned by the
configuration vectors $|n\rangle$ is usually easily expressed in
terms of bosonic creation and annihilation operators
$a^{\dagger}_i,a_i$ ($[a_i,a^{\dagger}_j]=\delta_{ij}$,
$[a_i,a_j]=0$). This is true for reaction-diffusion processes involving {\it bosonic
particles}, that is, without
exclusion, for which bosonic operators are particularly well-suited,
but this is also true when particles exclude each other. Nevertheless
in the latter situation
one should not expect that $\hat{H}$ will be a polynomial in terms of
the $a_i,a_j^{\dagger}$, while it indeed is in the former.\\

We confine the subsequent analysis to the dynamics of hard-core ({\it i.e.} mutually excluding
particles) undergoing diffusion and reaction processes.
\subsection{Evolution operators for some elementary processes}
In order to deal with the exclusion constraint we introduce the operator
$\delta_{\hat{n},m}$ defined by 
\begin{equation}\delta_{\hat{n},m}|m'\rangle=\delta_{m',m}|m'\rangle
\end{equation}
where $|m\rangle$ denotes a single site state. We now give two examples. The evolution operator for one-dimensional asymmetric diffusion in the presence of hard-core
interactions reads
\begin{equation}\label{opdiff}
\hat{H}_{\text{diff}}=\sum_i\left[
(D+\frac{v}{2})(1-a_i a^{\dagger}_{i+1})\delta_{\hat{n}_i,1}\delta_{\hat{n}_{i+1},0}
+(D-\frac{v}{2})(1-a_i^{\dagger} a_{i+1})\delta_{\hat{n}_{i},0}\delta_{\hat{n}_{i+1},1}\right]
\end{equation}
where $D+\frac{v}{2}$ (resp. $D-\frac{v}{2}$) is the hopping rate to the right
(resp. to the left).\\

For the simple $A+A\stackrel{k}{\to}\emptyset$ annihilation reaction of nearest
neighbor particles, one finds
\begin{equation}
\hat{H}_k=k\sum_i\left[
(1-a_i^{\dagger} a_{i+1}^{\dagger})\delta_{\hat{n}_i,1}\delta_{\hat{n}_{i+1},1}
\right]
\label{ann}\end{equation}
Extension to two-species annihilation is straighforward since bosonic operators
pertaining to distinct species commute.
\subsection{Passing to a coherent state representation}
In order to pass to a path-integral formulation, it is sufficient to follow the
steps described in \cite{cardytauber} (for a thorough and pedagogical
introduction we refer the reader to the review by Mattis and
Glasser~\cite{mattisglasser}). The result of those steps can be summarized as
follows. There exists an action $S[\hat{\phi},\phi]$ such that physical
observables can be expressed as path-integrals over the complex fields
$\hat{\phi}_i(t),\phi_i(t)$ of functions of those fields, weighted by
$\exp(-S)$. We denote by $[0,t_f]$ the time interval over which the process is
studied. The action $S$ has the form
\begin{equation}
S=-\sum_i\phi_i(t_f)+\int_0^{t_f}\dd t\;(\sum_i\hat{\phi}_i(t)\p_t\phi_i+H[\hat{\phi},\phi])
\end{equation}
where
\begin{equation}
H[\hat{\phi},\phi]=\frac{\langle\phi|\hat{H}|\phi\rangle}{\langle\phi|\phi\rangle}
\end{equation}
in which the notation $|\phi\rangle=\otimes_i|\phi_i(t)\rangle$ denotes the
tensor product of the coherent states
associated to each creator and annihilator $a^{\dagger}_i,a_i$ with eigenvalue
$\phi_i(t)$ (and $\hat{\phi}_i(t)$ denotes the complex conjugate of
$\phi_i(t)$). In order to evaluate the quantity $\frac{\langle\phi|\hat{H}|\phi\rangle}{\langle\phi|\phi\rangle}$, one normal
orders $\hat{H}$ and then simply replaces the $a_i$'s by $\phi_i(t)$ and the
$a_i^{\dagger}$'s by $\hat{\phi}_i(t)$. Hence the only difficulty is to be able
to normal order such an operator as $\delta_{\hat{n},m}$ (possibly multiplied by
$a$'s or $a^{\dagger}$'s). This can be done rather easily in several ways. One
of them is simply to normal order $\ee^{iu\hat{n}}$ then make use of the
integral representation 
$$\delta_{\hat{n},m}=\int_{-\pi}^{\pi}\frac{\dd
u}{2\pi}\ee^{iu(\hat{n}-m)}$$
Normal ordering $\ee^{iu\hat{n}}$ is done by expanding the exponential and
looking at each term in the series. It is amusing to
note that
\begin{equation}\label{ex1}
:\ee^{iu\hat{n}}:=\sum_{\ell=0}^{\infty}\frac{(iu)^\ell}{\ell!}\sum_{j=1}^\ell
s_{j,\ell}a^{\dagger j}a^j
\end{equation}
where the coefficients $s_{j,\ell}$ are the Stirling numbers of the second kind
($s_{j,\ell}=\frac{1}{n!}\frac{\dd^n}{\dd x^n}(\ee^{x}-1)^j\Big|_{x=0}$ is the number of ways a set with
$\ell$ elements can be partitioned into $j$
disjoint, non-empty subsets). Similarly, one finds that
\begin{equation}\label{ex2}
:\ee^{iu a^{\dagger}a}a^{\dagger}:=a^{\dagger}:\ee^{iu
a^{\dagger}a}:+\sum_n\frac{(iu)^n}{n!}\sum_j j s_{j,n} a^{\dagger j}a^{j-1}
\end{equation}
Once in their normal-ordered form, the operators $a$ and $a^{\dagger}$ in
Eqs~.(\ref{ex1}) and (\ref{ex2}) can be replaced by their coherent state eigenvalues. For
instance,
\begin{equation}
\langle\phi|\ee^{iu\hat{n}}|\phi\rangle
=\sum_{\ell=0}^{+\infty}\frac{(iu)^\ell}{\ell!}\sum_{j=1}^{\ell}s_{j,\ell}\hat{\phi}^j\phi^j
=\ee^{\hat{\phi}\phi(\ee^{iu}-1)}
\end{equation}
so that
\begin{equation}
\langle\phi|\delta_{\hat{n},0}|\phi\rangle=\int_{-\pi}^{\pi}\frac{\dd
u}{2\pi}\ee^{\hat{\phi}\phi(\ee^{iu}-1)}=\ee^{-\hat{\phi}\phi}
\end{equation}
Similar manipulations lead to the dictionary
\begin{eqnarray}
\langle\phi|a^\dagger\delta_{\hat{n},m}|\phi\rangle =
\frac{1}{m!}\hat{\phi}(\hat{\phi}\phi)^m\ee^{-\hat{\phi}\phi}\nonumber\\
 \langle\phi|\delta_{\hat{n},m}
 |\phi\rangle=\frac{1}{m!}(\hat{\phi}\phi)^m\ee^{-\hat{\phi}\phi}\nonumber\\
 \langle\phi|a\delta_{\hat{n},m}|\phi\rangle 
 =\phi\frac{1}{(m-1)!}(\hat{\phi}\phi)^{m-1}\ee^{-\hat{\phi}\phi}\;\;(=
 0\text{ if }m=0)\label{dico}
\end{eqnarray}
In a reaction-diffusion process with hard-core particles, only the formulas for
$m=0$ or $1$ will be needed. Formulas with higher powers of the $a$'s or
$a^{\dagger}$'s are derived in a similar way.\\

For example, the action for the annihilation process described by the evolution
operator of Eq.~(\ref{ann}) reads
\begin{equation}
S=k\int\dd
t\sum_i\left[(\hat{\phi}_i\hat{\phi}_{i+1}-1)\phi_i\phi_{i+1}\ee^{-\hat{\phi}_{i}\phi_{i}-\hat{\phi}_{i+1}\phi_{i+1}}\right]
\end{equation}
Extensions to several species
poses no new problem. A final remark: when studying a process that conserves
the parity of the instantaneous total number of particles $N(t)$ in the system, the
following quantity is conserved
\begin{equation}
\langle(-1)^{N(t)} \rangle=\langle\ee^{i\pi N(t)} \rangle=\langle\text{\bf
p}|\ee^{i\pi\sum_{i}\hat{n}_i}|\Psi\rangle
\end{equation}
where $\langle \text{\bf p}|\equiv\langle 0|\ee^{\sum_i a_i}$ denotes the projection
state~\cite{cardytauber}, which is also a coherent state with eigenvalue 1, so
that, using that $\langle 1|\ee^{i\pi
\hat{n}_i}|\phi_i\rangle=\ee^{\phi_i(\ee^{i\pi}-1)}$, we find
\begin{equation}
\langle(-1)^{N(t)} \rangle=\langle\ee^{-2\sum_{i}\phi_i(t)}\rangle=\text{cst}
\end{equation}
This is a way of characterizing parity by means of a
well-defined observable. We refer the reader to Deloubri\`ere and
Hilhorst~\cite{olivierhenk} for further comments in the context of the pair
annihilation reaction.
\subsection{The Park, Kim and Park approach}
It is well-known~\cite{gardiner} that for reaction-diffusion processes involving bosonic particles (which is not the case in \cite{scpark}) it is
possible to write a partial differential equation for some continuous random
variable $\rho_i$ (the mapping fails for hard-core particles). The first moment of $\rho_i$ equals the average local particle
number $\langle n_i\rangle$ (which makes it tempting to identify $\rho_i$ with a
fluctuating density). However higher moments of $\rho_i$
do not coincide with those of $n_i$ (though they can be related). This partial
differential
equation takes a Fokker-Planck form ({\it i.e.} is of order 2) only when
the microscopic reactions involve at most two particles.\\

We now refer to the article \cite{scpark} in which the authors have presented an
alternative route to derive a path-integral formulation for the dynamics of hard-core
particle systems. Particle numbers are discrete variables, so that there is no
Fokker-Planck equation for them. In their Eq.~(7) they write a Fokker-Planck
equation for some continuous random variable $\rho_i$, which, as we have said, is
not correct without further approximation. This error is not related to the necessity of
implementing the hard-core constraint or not. Such a Fokker-Planck equation
simply does not exist.\\

References and further comments can be found in the book by
Gardiner~\cite{gardiner} or in Deloubri\`ere and Hilhorst~\cite{olivierhenk}.\\

\section{An example: Asymmetric diffusion of hard-core particles and the noisy
Burgers equation}
As an example we recover the noisy Burgers equation by going to the continuous limit in the
asymmetric diffusion of a system of hard-core particles. While this equivalence
is certainly not new~\cite{nelson,beijeren,janssenschmittmann}, we use it as a testbench for the method. It should be
mentioned that this derivation of the noisy Burgers equation is the first one
that starts from an exact mapping. 
\subsection{Action}
Again, for notational simplicity, we restrict the analysis to one space dimension. We
consider diffusion with a hopping rate to the right $D+\frac{v}{2}$ and a
hopping rate to the left $D-\frac{v}{2}$. The evolution operator for such asymmetric diffusion in the presence of hard-core
interactions is that of Eq.~(\ref{opdiff}). Using the dictionary Eq.~(\ref{dico}) we find that the corresponding action reads
\begin{equation}\begin{split}
S[\hat{\phi},\phi]=\int\dd
t\sum_i\Bigg[&\hat{\phi}_i\p_t\phi_i\\
&+(D+\frac{v}{2})
(\hat{\phi}_{i}-\hat{\phi}_{i+1})\phi_i
\ee^{-\hat{\phi}_{i}\phi_{i}-\hat{\phi}_{i+1}\phi_{i+1}}\\&
+(D-\frac{v}{2})
(\hat{\phi}_{i}-\hat{\phi}_{i-1})\phi_i
\ee^{-\hat{\phi}_{i}\phi_{i}-\hat{\phi}_{i-1}\phi_{i-1}}\Bigg]
\end{split}\label{adiff}\end{equation}
No
approximation was made and the action Eq.~(\ref{adiff}) is fully exact. Without
the exponential factors Eq.~(\ref{adiff}) would
yield the usual action of asymmetric diffusion for bosonic particles.
Here, owing to the presence of the nonlinear interaction terms (the exponentials), there is no Galilean
transformation that eliminates the drift-dependent terms. 

\subsection{Recovering the noisy Burgers equation in the continuous limit}
In this paragraph, we shall show that expanding naively the action Eq.~(\ref{adiff}) leads to the
noisy 
Burgers equation for the density fluctuations, as it should. We perform the
change of fields~\cite{janssencardy}
${\phi}_i=(\rho+\psi_i)\ee^{-\bar{\psi}_i}$,
$\hat{\phi}_i=\ee^{\bar{\psi}_i}$. Now we expand the action in powers of the new
fields $\bar{\psi}_i$ and $\psi_i$ (the latter represents a fluctuation of the density with respect to its
average value $\rho$). We also assume that the fields have slow space variations,
and take the limit of a continuous space. The resulting action reads
\begin{equation}
S=\int\dd t\dd x\;\left[\bar{\psi}(\p_t+v\p_x-D\p_x^2)\psi-g_1(\p_x\bar{\psi})^2-g_2\bar{\psi}\psi\p_x\psi+... \right]
\label{burgers}\end{equation}
where the constants $g_1,g_2$ are positive functions of the microscopic
details of the model (diffusion constant, drift velocity, lattice
spacing, average density). The dots stand for higher polynomial or
higher derivative terms. The action Eq.~(\ref{burgers}) for the fields $\bar{\psi},\psi$
is equivalent to $\psi(x,t)$ satisfying the noisy Burgers equation~\cite{janssenschmittmann}. Hence, up to terms which are irrelevant in the scaling
limit (see Janssen and Schmittmann~\cite{janssenschmittmann} for a
renormalization group analysis of the action Eq.~(\ref{burgers})), we have
recovered the equivalence between asymmetric diffusion of hard-core particles
and the noisy Burgers equation. As a final remark, we would like to emphasize
that we have not resorted to any phenomenological arguments, and that our
derivation is completely systematic -- the first one of this sort. This feature
is particularly encouraging since we have in mind the application of the formalism to other less studied
processes.

\section{Conclusions}
We have shown how to build up a field-theoretic formalism that takes into account in a
systematic fashion the effect of exclusion in $d$-dimensional reaction-diffusion
processes. We have exemplified the formalism on the case of asymmetric
diffusion, thus recovering the noisy Burgers equation. We now have a tool to take up any reaction-diffusion process
in which one, or all species, diffuse with a drift, such as the $A+B\to
\emptyset$ reaction, for which it is conjectured that exclusion changes the
universality class of the scaling behavior. The $A+B\to\emptyset$
reaction-diffusion process with drift is certainly the first system that should be looked at using the present
approach. However, great care must be paid to naive expansions of
exponential interaction terms, and the feasibility of such a procedure must be investigated in
each particular case.\\

The method presented here opens the door to the study of reaction-diffusion processes in which exclusion
is conjectured to play a crucial r\^ole, such as in the $N$ species branching annihilating random walks
 recently described by Kwon {\it et
al.}~\cite{park}.\\

\noindent {\bf Acknowledgments:} The author would like to thank Henk Hilhorst,
Uwe T\"auber and Martin Howard without whom this approach would not have been
elaborated, and would like to thank Gunter Sch\"utz for communicating
\cite{schutz} prior to publication.

\end{document}